\documentclass[twocolumn,amsmath,amssymb,superscriptaddress,floatfix,showpacs]{revtex4}
\usepackage{color}
\usepackage{graphicx} 
\usepackage{tabularx} 
\usepackage{hyperref}

\newcommand{\vev}[1]{\langle{#1}\rangle}
 
\def\dsl{\partial\hspace{-2.1mm}/} 
\def\chis{\chi_\sigma}

 \graphicspath{ 
 {./} 
 {./../figures/}
 {./../figures/phase_diagram_cutoff/} 
 {./../output/}
 {../../mean_field_quark/output/}
 }
    
\begin{document}
\title{\bf Susceptibilities near the QCD (tri)critical point}
\author{B.-J. Schaefer}
\email[E-Mail:]{bernd-jochen.schaefer@uni-graz.at}
\affiliation{Institut f\"{u}r Physik, Karl-Franzens-Universit\"{a}t,
  A-8010 Graz, Austria}  
\affiliation{Institut f\"{u}r Kernphysik, TU Darmstadt, D-64289
  Darmstadt, Germany} 
\author{J. Wambach} \affiliation{Institut
  f\"{u}r Kernphysik, TU Darmstadt, D-64289 Darmstadt, Germany}
\affiliation{Gesellschaft f\"{u}r Schwerionenforschung GSI, D-64291
  Darmstadt, Germany}

 
\date{\today}
 
\begin{abstract}
  Based on the proper-time renormalization group approach, the scalar
  and the quark number susceptibilities in the vicinity of possible
  critical end points of the hadronic phase diagram are investigated
  in the two-flavor quark-meson model. After discussing the quark-mass
  dependence of the location of such points, the critical behavior of
  the in-medium meson masses and quark number density are calculated. The
  universality classes of the end points are determined by calculating
  the critical exponents of the susceptibilities.  In order to
  numerically estimate the influence of fluctuations we compare all
  quantities with results from a mean-field approximation. It is
  concluded that the region in the phase diagram where the
  susceptibilities are enhanced is more compressed around the critical
  end point if fluctuations are included.
\end{abstract}

\maketitle
\section{Introduction}

Theoretical studies reveal an increasing richness in the structure of
the phase diagram of strongly interacting matter. At high temperature
$T$ and/or quark chemical potential $\mu$ it is expected that the
system undergoes a phase transition from the ordinary hadronic
phase to a chirally restored and deconfined quark gluon plasma
(QGP)~\cite{Stephanov1999, Stephanov2005}. Other phases such as a
two-flavor (2SC) and color-flavor locked (CFL) color-superconducting
phase are predicted at high densities and small temperatures. For
small chemical potential and high temperature recent lattice
calculations suggest the existence of mesonic bound states even above
the deconfinement phase transition temperature~\cite{Datta2004},
indicating a strongly coupled QGP.

Based on model calculations~\cite{Asakawa1989, Barducci1989,
  Barducci1990, Berges1999a} as well as lattice QCD
simulations~\cite{Fodor2002, Forcrand2002, Allton2003}, the existence
of a critical endpoint (CEP) )in the phase diagram is suggested. This
is the endpoint of a first-order transition line in the
$(T,\mu)$-plane and is a genuine singularity of the QCD free energy.
Here the phase transition is of second order, belonging to the
three-dimensional Ising universality class. Its precise location,
which is e.g.~highly sensitive to the value of the strange-quark mass,
is not known at present. It might be accessible with current and
future experimental facilities and its observable implications in
relativistic heavy-ion experiments such as event-by-event fluctuations
of suitable observables are intensively
discussed~\cite{Stephanov2004}.

From lattice studies it is known that the dynamics of the transition
along the temperature axis is strongly affected by the presence of
light quarks. The transition temperature is lowered substantially from
its value in the pure gauge limit of infinitely heavy quarks, mostly
due to the nearly massless up- and down quarks. This indicates a
prominent role of the (nearly exact) chiral $SU(2)_L \times SU(2)_R$
symmetry, which is spontaneously broken to $SU(2)_V$ at small
temperatures and baryon densities. Indeed, a chiral phase transition
is observed numerically whose transition temperature coincides with
that for the deconfinement transition. In the limit of vanishing up-
and down quark masses and infinite strange quark mass, the chiral
phase transition is likely to be of second-order at vanishing $\mu$
and the static critical behavior is expected to fall in the
universality class of the Heisenberg $O(4)$ model in three
dimensions~\cite{Pisarski1984}.

Lattice calculation of the QCD phase transition with finite $\mu$ is
much more difficult. Due to the Fermion sign problem, simulations of
similar precision as at $\mu=0$ are still lacking. From direct
numerical evaluation of the QCD partition function and its
quantum-statistical analysis~\cite{Fodor2002, Forcrand2002} or from a
Taylor expansion of the pressure around $\mu=0$ there is, however,
evidence for a CEP at finite $\mu$. At this point both the chiral- and
the quark number susceptibility diverge. The existence of a CEP in the
($T,\mu$)-plane implies that massless two-flavor QCD has a tricritical
point (TCP) at which the $O(4)$ line of critical points ends. For
larger $\mu$'s the transition is then of first-order.

In an attempt to interprete the physical content of the lattice
results there is a variety of model studies of the CEP and the
critical region around it~\cite{Barducci1994, Halasz1998, Harada1999,
  Hatta2003, Fujii2003, Brouzakis:2004se}. In future searches for the
CEP in heavy-ion reactions the size of the critical region is
especially important~\cite{Nonaka2005}. Most of these studies rely in
one way or another on a mean-field description of the phase
transition. As is well known, mean-field theory very often fails to
give the correct description of phase transitions and fluctuations
have to be included. An efficient way to describe critical phenomena
beyond mean-field theory is the renormalization group (RG) method. It
can be used to describe the universal and non-universal aspects of
second-order as well as first-order phase transitions. In the context
of the phase diagram of strongly interacting matter the RG-method has
been applied in~\cite{Schaefer2005} using a two-flavor quark-meson
model, which captures essential chiral aspects of QCD.

Extending our previous analysis, the present work primarily focuses
is the question how large the critical regions around the TCP and CEP
might be. The size of a critical region is defined as the one where
mean-field theory breaks down and non-trivial critical exponents
emerge.  A well-known criterion is the Ginzburg criterion which
estimates the size by some unknown coupling coefficients. It is based
on an expansion of the singular part of the free energy for a
second-order phase transition. Since the expansion coefficients,
appearing in this criterion are not known for strong interactions the
Ginzburg criterion is of limited use in the present context. Also
universality arguments are not helpful, if the underlying microscopic
dynamics is not well determined. This can be seen, for example, in the
$\lambda$ transition of liquid helium He$^4$ and the superconducting
transition of metals. Both transitions belong to the same universality
class of the $O(2)$ spin model but their critical regions defined by
their corresponding Ginzburg-Levanyuk temperature $\tau_{GL}$, differ
by several orders of magnitude. For hadronic matter it is expected
that the critical region of the CEP is small~\cite{Hatta2003}.

The paper is organized as follows. In the next
Sec.~\ref{Sec:meanfield} a mean-field analysis of the quark-meson
model for two quark flavors as an effective realization of the
low-energy sector of QCD is presented.  After the derivation of the
grand canonical potential for the model, we calculate the phase
diagram and localize the CEP. The behavior of the scalar- and
pseudoscalar meson masses, the quark number density and the scalar and
quark number susceptibilities near the CEP are investigated. In order
to obtain a deeper understanding of the shape of the critical region
around the CEP, critical exponents are calculated. In
Sec.~\ref{Sec:PTRG} we repeat the calculation using the proper-time
renormalization group approach (PTRG) in order to assess the influence
of fluctuations.  Finally, before summarizing in Sec.~\ref{Sec:sum},
the size of the critical regions around the CEP and TCP are determined
and compared with the mean-field results.

\section{The quark-meson model}

The Lagrangian of the linear quark-meson model for $N_f=2$ light
quarks $q=(u,d)$ and $N_c=3$ color degrees of freedom reads
\begin{eqnarray} \label{eq:qmmodel}
  {\cal L} &=& \bar{q} \,(i\dsl - g (\sigma + i \gamma_5
  \vec \tau \vec \pi ))\,q 
  \nonumber \\
&& \quad +\frac 1 2 (\partial_\mu \sigma \partial^\mu\sigma+
  \partial_\mu \vec \pi  \partial^\mu\vec \pi) - U(\sigma, \vec \pi )
\end{eqnarray}
where the purely mesonic potential is defined as
\begin{eqnarray} \label{eq:pot}
U(\sigma, \vec \pi ) &=& \frac \lambda 4 (\sigma^2+\vec \pi^2 -v^2)^2
-c\sigma\ .
\end{eqnarray} 
The isoscalar-scalar $\sigma$ field and the three
isovector-pseudoscalar pion fields $\vec \pi$ together form a chiral
4-component field $\vec \phi$. Without the explicit symmetry breaking term
$c$ in the potential the Lagrangian is invariant under global chiral
$SU(2)_L\times SU(2)_R$ rotations.

The four parameters of the model are chosen such that the chiral
symmetry is spontaneously broken in the vacuum and the $\sigma$ field
develops a finite expectation value $\vev \sigma \equiv f_\pi$, where
$f_\pi=93$ MeV is the pion decay constant. Due to the
pseudoscalar character of the pions the corresponding expectation
values vanish $\vev {\vec \pi} =0$.

The Yukawa coupling constant $g$ is usually fixed by the constituent
quark mass in the vacuum $g= M_q/f_\pi$. Using the partially conserved
axial vector current (PCAC) relation, the explicit symmetry breaking
parameter $c$ is determined by $c=M_\pi^2 f_{\pi}$, where $M_{\pi}$ is
the pion mass. The quartic coupling constant $\lambda$ is given by the
sigma mass $M_{\sigma}$ via the relation $\lambda= \frac 1 {2f_\pi^2}
(M_\sigma^2 -M_\pi^2)$. Finally, the parameter $v^2$ is found by
minimizing the potential in radial direction, yielding $v^2 =
\sigma^2-c/(\lambda \sigma)$. For the ground state where $\vev \sigma=
f_{\pi}$ this expression can be rewritten as $v^2 = f_{\pi}^2
-M^2_\pi/\lambda$ .  It is positive in the Nambu-Goldstone phase.

\section{Mean-field approximation} 
\label{Sec:meanfield}

\subsection{The thermodynamic potential}

In a spatially uniform system the grand canonical potential (per
volume) $\Omega$ is a function of the temperature $T$ and of the quark
chemical potential $\mu$. We confine ourselves to the
$SU(2)_f$-symmetric case and set $\mu \equiv \mu_u = \mu_d$. This is a
good approximation since flavor mixing between the $u$- and $d$-quark
in the vector channel is very small.

The grand canonical potential is obtained as the logarithm of the
partition function, which is given by a path-integral over the meson
and quark/antiquark fields
\begin{eqnarray} \label{eq:path} 
{\cal Z} \!\!&=& \!\!\!\int \!\!{\cal D} \bar{q}  {\cal D} q {\cal D} \sigma {\cal
  D} \vec{\pi} \exp \left\{ \!\!\int\limits_0^{1/T} dt d^3x \left( {\cal L} +\mu
  \bar{q} \gamma_0 q \right)\right\}.
\end{eqnarray} 
To start with we evaluate the partition function in the mean-field
approximation similar to Ref.~\cite{Scavenius2001}. Thus we replace the
meson fields in Eq.~(\ref{eq:path}) by their expectation values in the
action neglecting the quantum and thermal fluctuations of
the mesons. The quarks/antiquarks are retained as quantum fields. The
integration over the fermions yields a determinant which can be
calculated by standard methods (see e.g.~Ref.~\cite{Kapusta1989}).
This generates an effective potential for the mesons.  Finally, one
obtains 
\begin{eqnarray} \label{eq:totpot}
\Omega(T,\mu) &=& -\frac {T \ln {\cal Z} } V = U( \vev{\sigma} ,
\vev{\vec \pi}) + \Omega_{\bar{q} q} (T,\mu)
\end{eqnarray}
with the quark/antiquark contribution
\begin{eqnarray} \label{eq:qqcontribution}
\Omega_{\bar{q} q} (T,\mu)&\!\!\!=\!\!\!& \nu_q T\!\!
\int\limits\!\!\!\frac{d^3k}{(2\pi)^3} 
\left\{\ln(1-n_q(T,\mu))\right.\nonumber \\
&&\qquad\qquad\left. + \ln(1-n_{\bar{q}}(T,\mu))
\right\}
\end{eqnarray} 
where 
\begin{eqnarray}\label{eq:occ}
n_q(T,\mu) = \frac 1 {1+\exp((E_q-\mu)/T)} 
\end{eqnarray}
resp.~$n_{\bar q}(T,\mu)$ are the usual quark/antiquark occupation
numbers with $n_{\bar q}(T,\mu) \equiv n_q(T,-\mu)$. The
single-particle energy is given by $E_q = \sqrt{{\vec k}^2+ M_q^2}$
and the effective constituent quark mass by $M_q = g \vev \sigma$. The
number of internal quark degrees of freedom is denoted by $\nu_q=2N_c
N_f =12$.

The divergent vacuum contribution which results from the negative energy
states of the Dirac sea has been neglected in
Eq.~(\ref{eq:qqcontribution}) assuming that the renormalization is
already done in the vacuum. This omission of the vacuum contribution
is in contrast to a similar analysis in the Nambu--Jona-Lasinio model
where the vacuum contribution in the corresponding grand canonical
potential cannot be neglected due to the finite momentum cutoff
regularization~\cite{Scavenius2001}.

Furthermore, we remark here that the same quark/antiquark contribution
of the grand canonical potential can also be obtained by an analytical
integration of the corresponding RG flow equation in mean-field
approximation \cite{Schaefer2005}.  In this simple approximation
the mesonic RG flow is neglected and the residual quark/antiquark flow
is given by
\begin{eqnarray} 
\partial_k \Omega_{\bar q q} &=&  -\frac{\nu_q k^4}{12\pi^2}\frac 1
{E_q}
\left\{ \tanh(\frac{E_q-\mu}{2T}) \right.\nonumber \\
&&\hspace*{3cm} \left. + \tanh(\frac{E_q+\mu}{2T})
\right\}\nonumber \\
&=& -\frac{\nu_q k^4}{6\pi^2}\frac 1
{E_q}
\left[ 1 - n_q(T,\mu) - n_{\bar q}(T,\mu) \right],
\end{eqnarray} 
which, after the scale integration towards the infrared, is identical
to Eq.~(\ref{eq:qqcontribution}) if the vacuum contribution is neglected.

For the case of a massless free quark/antiquark gas
Eq.~(\ref{eq:qqcontribution}) can be integrated analytically with the
result
\begin{eqnarray} \label{eq:freepot}
\lim_{M_q\to 0}\Omega_{\bar{q} q} (T,\mu) = -\frac{\nu_q} 2 \left[
  \frac{\mu^4}{12\pi^2} +\frac{\mu^2 T^2}{6}
  +\frac{7\pi^2T^4}{180}\right].
\end{eqnarray} 

In general, any phase transition is characterized by an order
parameter $\phi$ which is identified here with the expectation value
of the sigma field. Its behavior is determined by the corresponding
classical equation of motion. It is obtained by minimizing the
potential $\Omega$ in radial $\sigma$-direction. This results in the
gap equation
\begin{eqnarray}\label{eq:gap}
\frac {\partial \Omega}{\partial \sigma} &=& 0 \ =\ \lambda\vev \sigma
(\vev\sigma^2 -v^2) - c + g \vev{\bar qq}
\end{eqnarray}
where the scalar quark density $\vev{\bar qq}$ has been introduced
\begin{eqnarray} 
\vev{\bar qq} &=& g \vev\sigma \nu_q  \int\!\!\!\frac{d^3k}{(2\pi)^3}
\frac 1 {E_q}\left\{n_q(T,\mu) + n_{\bar q}(T,\mu) \right\}\ .
\end{eqnarray} 
\footnote{In order to compare the resulting expressions directly with
  the RG treatment to be discussed later we consider derivatives
  w.r.t.~the squared field $\sigma^2$, which has numerical
  advantages.}
 
The quark density represents the source term in the equation of motion
and can also be calculated analytically for vanishing masses.  In this
limit, one obtains for the quark density
\begin{eqnarray}\label{eq:limqq}
\lim_{M_q\to 0} \frac{\vev{\bar qq}}{M_q} &=& {\nu_q} \left[ \frac {T^2}{12} +
  \frac{\mu^2}{4\pi^2}\right]\ .
\end{eqnarray} 

\subsection{The phase diagram}

The solution of the gap equation (\ref{eq:gap}) determines the
behavior of the order parameter as a function of the temperature and
the quark chemical potential and thus allows to study the phase
structure of the underlying two-flavor quark-meson model.

In the vacuum we fix the model parameters to $M_{\pi} =138$ MeV,
$M_{\sigma} =600$ MeV, $f_{\pi} =93$ MeV and $M_q = 300$ MeV which
result in $c \sim 1.77\cdot 10^6$ MeV$^3$, $v\sim 87.6$ MeV, $\lambda
\sim 19.7$ and $g \sim 3.2$.

\begin{figure}[!htb]
  \centerline{\hbox{
      \includegraphics[width=0.6\columnwidth,angle=-90]{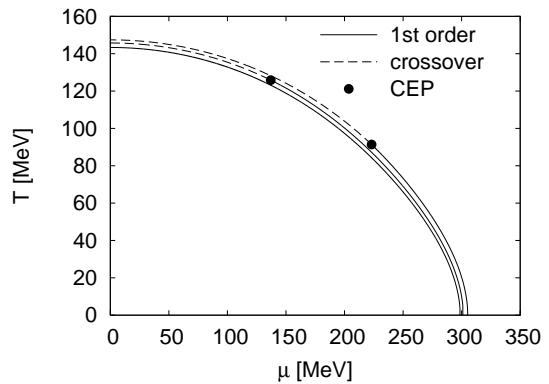}
    }}
  \caption{\label{fig:mfphasediagram} The phase diagram of the linear
    quark-meson model with two constituent quarks in mean-field
    approximation for three different fits of the pion mass.  From top
    to bottom: $M_{\pi} =138,100,69$ MeV.}
\end{figure}

For a physical pion mass ($M_{\pi} =138$ MeV) the model exhibits a
smooth crossover on the temperature axis and a first-order chiral
phase transition on the density axis. For increasing temperatures this
first-order transition line persists up to a critical endpoint (CEP)
where the chiral transition becomes second-order.
Along the line of a first-order phase transition the thermodynamic
potential has two minima of equal depth. These minima are separated by
a finite potential barrier making the potential non-convex, which is
typical for a mean-field approximation.  The height of the barrier is
largest at zero temperature and finite quark chemical potential and
decreases towards higher temperature. At the critical endpoint the
barrier disappears (no latent heat) and the potential flattens.  For
temperatures above the CEP the transition is washed out and a smooth
crossover takes place. With the parameters chosen above the location of the 
CEP is found to be at $T_c = 91.4 \mbox{ MeV},~\mu_c = 223 \mbox{ MeV}$.
In Fig.~\ref{fig:mfphasediagram} the phase diagram in the
complete $(T,\mu)$-plane is shown.  

To study the influence of explicit chiral symmetry breaking we have
varied the model parameters keeping the Yukawa coupling $g$ fixed. The
resulting phase boundaries are also shown in
Fig.~\ref{fig:mfphasediagram} for $M_\pi =100$ MeV and $M_\pi = 69$
MeV (half of the physical pion mass).  Reducing the pion mass while
keeping the Yukawa coupling fixed, the CEP moves towards the $T$-axis.
This is also the case if the Yukawa coupling is increased and the pion
mass is kept fixed.  Already for the pion mass $M_\pi = 69$ MeV the
CEP disappears and chiral symmetry is restored via a first-order
transition for all temperatures and quark chemical potentials. As a
consequence this model does not have a tricritical point in the chiral
limit (see also App.C in~\cite{Fujii2004}).  This is in contradiction
to universality arguments as well as lattice QCD simulations.  At
vanishing quark chemical potential, the effective theory for the
chiral order parameter is the same as for the $O(4)$ model, which has
a second-order phase transition \cite{Kapusta1989}.  It is further
expected that the static critical behavior falls into the universality
class of the $O(4)$-symmetric Heisenberg model in three
dimensions~\cite{Pisarski1984}.

But, as already stated in Ref.~\cite{Mocsy2002}, within the mean-field
approximation the order of the phase transition in the chiral limit of
the linear quark-meson model strongly depends on the values for the
model parameters. The way how to extrapolate towards the chiral limit
is not unique. As we shall see below, the mean-field approximation
itself is also questionable~\cite{Baym1977,Aouissat1996}.

\begin{figure}[!htb]
  \centerline{\hbox{
      \includegraphics[width=0.6\columnwidth,angle=-90]{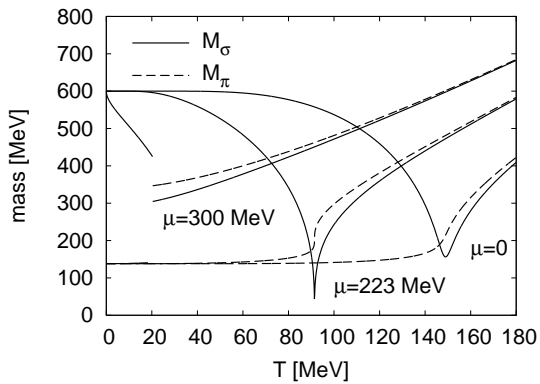}
    }}
\caption{\label{fig:masses_T} The pion and sigma masses as
  function of the temperature for three different quark chemical
  potentials ($\mu=0$, $\mu=\mu_c\sim223$ and $\mu=300$ MeV). In the
  vicinity of the CEP (at $\mu_c$ and $T_c \sim 91$ MeV) the sigma
  mass $M_\sigma$ vanishes.}
\end{figure}

\begin{figure}[!htb]
  \centerline{\hbox{
      \includegraphics[width=0.6\columnwidth,angle=-90]{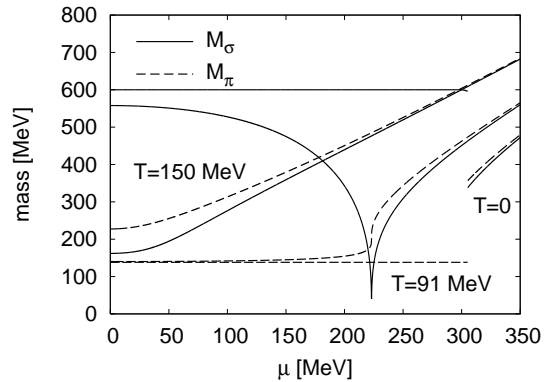}
    }}
\caption{\label{fig:masses_mu} Similar to Fig.~\ref{fig:masses_T}: The
  pion and sigma masses as function of the chemical potential for
  three different temperatures ($T=0$, $T= T_c\sim 91$ and $T=150$
  MeV).}
\end{figure}

\subsection{In-medium meson masses}

To gain further physical insight into the critical behavior of the
model we have studied the 'in-medium' meson masses $M_\pi(T,\mu)$ and
$M_\sigma (T,\mu)$, which encode the critical fluctuations of the
matter. In Fig.~\ref{fig:masses_T} the temperature dependence of the
meson masses for three different chemical potentials are shown. The
sigma mass always decreases with temperature in the hadronic (chirally
broken) phase and increases again at high temperature. The pion mass
does not vary much below the transition but increases at high
temperatures similar to the sigma mass indicating the restoration of
chiral symmetry. The increase of the sigma and pion mass at high
temperature is driven by the asymptotic behavior of the scalar quark
density (see Eq.~(\ref{eq:limqq})) because chiral symmetry restoration
enforces a decrease of the constituent quark mass. At large
temperature the meson masses degenerate and increase linearly with
$T$. A similar behavior is seen for the corresponding screening masses
on the lattice at $\mu=0$~\cite{Karsch2002}. For very high $T\gg T_c$
the meson screening masses approach $2\pi T$ because they are
essentially controlled by the lowest Matsubara frequency of the quarks
\cite{Eletsky1988} \cite{Gocksch1991}. But for temperatures slightly
above $T_c$ the screening masses in the scalar and pseudoscalar
channels are both still below $2\pi T$, which indicates a rather
strong $q$-$\bar{q}$ interaction in the high-$T$ phase.

Fig.~\ref{fig:masses_mu} shows the chemical potential dependence of
the meson masses for three different temperatures analogous to
Fig.~\ref{fig:masses_T}. For small temperatures a discontinuity is
visible in the evolution of the masses, signaling a first-order phase
transition. At $T=0$ and small $\mu$ the masses are constant and equal
to the vacuum masses which can also be seen analytically. At the CEP
the phase transition is of second order. Second-order phase
transitions are characterized by long-wavelength fluctuations of the
order parameter. Since the order parameter is proportional to the
scalar $\sigma$-field, the corresponding scalar sigma mass must vanish
at this point. In the vicinity of this point the sigma mass drops
below the pion mass and the potential flattens in the radial
direction. This behavior can clearly be seen in both
Figs.~\ref{fig:masses_T} and \ref{fig:masses_mu}. Since chiral
symmetry is still explicitly broken the pseudoscalar mass stays always
finite. For temperatures above the chiral transition the sigma mass
drops below the pion mass and increases with the chemical potential
(cf.~Fig.~\ref{fig:masses_T} and \ref{fig:masses_mu}). Close to
$\mu_c$ for $\mu=300$ MeV in Fig.~\ref{fig:masses_T} the sigma mass
drops rapidly with temperature and jumps at the chiral transition.

\subsection{Susceptibilities}

In order to find a bound to the size of the critical region around the
CEP we calculate the quark number susceptibility $\chi_q$ and its
critical behavior.  In general, the quark number susceptibility is the
response of the quark number density $n(T, \mu)$ to an infinitesimal
variation of the quark chemical potential
\begin{equation}\label{eq:chiq}
\chi_q (T, \mu) = \frac {\partial n(T,\mu)}{\partial\mu}\ .
\end{equation} 
Before we study the thermodynamic properties of the susceptibility,
 we start with an analysis of the (total) quark number
density $n(T,\mu)$. It is defined by a derivative w.r.t. the quark
chemical potential of the grand canonical potential (\ref{eq:totpot})
\begin{eqnarray}\label{eq:nqmf} 
n(T,\mu )\!\!&=&\!\! \vev{q^\dagger q}=
-\frac{\partial \Omega (T, \mu)} {\partial \mu}\nonumber\\
&=&\!\!\nu_q\!\int\!\!\!\frac{d^3k}{(2\pi)^3}\left\{ n_q(T,\mu ) -
  n_{\bar{q}}(T, \mu) \right\}.
\end{eqnarray} 

For $T\to 0$ and $\mu \to 0$ respectively, the analytical results for
the quark number density are
\begin{eqnarray} 
n(0,\mu) &=& \frac{\nu_q}{6\pi^2} (\mu^2-M_q^2)^{3/2}\
\theta(\mu-M_q),\\
n(T, 0) &=& 0\ .
\end{eqnarray}
For a massless free quark gas one obtains for the density
\begin{eqnarray} \label{eq:freedensity}
\lim_{M_q \to 0} n(T,\mu) &=& \frac{\nu_q} 6 \mu \left[ T^2 +
  \frac{\mu^2}{\pi^2}\right].
\end{eqnarray}

\begin{figure}[!htb]
  \centerline{\hbox{
      \includegraphics[width=0.6\columnwidth,angle=-90]{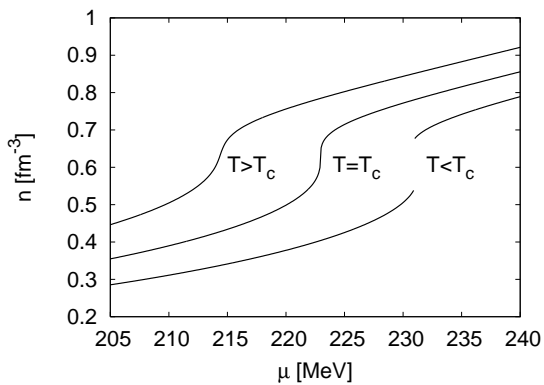} }}
  \caption{\label{fig:nq_cep_3T} The quark number density $n$ in
    mean-field theory as a function of the chemical potential $\mu$
    around the CEP for three different temperatures. The temperatures
    are $T_c \sim 91$ MeV and $T = T_c \pm 5$ MeV.  }
\end{figure} 

\begin{figure}[!htb]
  \centerline{\hbox{
      \includegraphics[width=0.6\columnwidth,angle=-90]{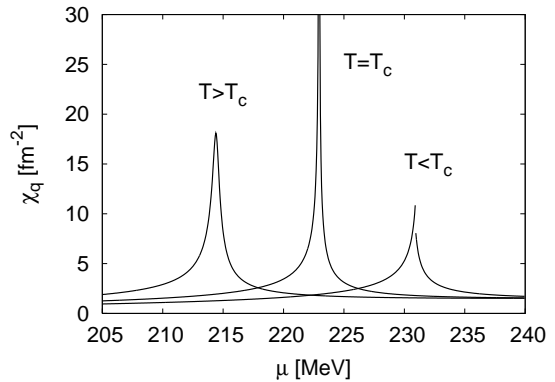}
    }}
  \caption{\label{fig:qsuscept_CEP_3T} The quark number susceptibility
    in mean-field approximation as a function of the chemical
    potential around the CEP for three temperatures. The temperatures
    are the same as in Fig.~\ref{fig:nq_cep_3T}. (See text for
    details).}
\end{figure}

In Fig.~\ref{fig:nq_cep_3T} the quark number density in mean-field
approximation is shown for three different temperatures around the
CEP. For $T=T_c$ the slope tends to infinity at $\mu=\mu_c$ which will
yield a diverging susceptibility.  For temperatures below the critical
one the system undergoes a first-order phase transition and the
quark number density jumps. The slope of the quark density in the
chirally symmetric phase is almost constant for all temperatures
around the CEP. For temperatures above the CEP the discontinuity
vanishes at the transition and the density changes gradually due to
the smooth crossover. This produces a finite height of the
quark number susceptibility $\chi_q$ which is calculated via
Eq.~(\ref{eq:chiq}). The resulting susceptibilities are shown in
Fig.~\ref{fig:qsuscept_CEP_3T}.

For $\mu=0$ the susceptibility $\chi_q (T,0)$ increases as the quark
mass $M_q(T)$ decreases and reaches $N_f T^2$ at $M_q(T)=0$ for three
colors. $\chi_q$ diverges only at the CEP and is
finite everywhere else. The height of $\chi_q$ decreases for
decreasing chemical potentials above the CEP towards the $T$-axis. For
temperatures below the CEP $\chi_q$ is discontinuous and jumps across
the first-order transition line.

It is instructive to note that the quark number susceptibility is
proportional to the isothermal compressibility $\kappa_T$ via the
relation
\begin{equation}
\kappa_T = \chi_q(T,\mu)/n^2(T,\mu) \ .
\end{equation}
In the vicinity of the critical point the quark number density $n$ is
always finite but the susceptibility becomes large. This indicates
that the system is easy to compress around the critical point and
suggests that the interaction between all constituents becomes very
attractive~\cite{Hatsuda:1994pi}.

It is also possible to express the quark number susceptibility as an
integral over the quark/antiquark occupation numbers (\ref{eq:occ})
\begin{eqnarray} 
\chi_q (T, \mu)&=& \frac{\nu_q}{T}\!\int\!\!\!\frac{d^3k}{(2\pi)^3}\left\{
  n_q(T,\mu ) (1-n_q(T,\mu ))\ + \right.\nonumber \\
&&\hspace*{5em} \left.n_{\bar{q}}(T, \mu)(1- n_{\bar{q}}(T,
  \mu))\right\}
\end{eqnarray} 
with the corresponding limits
\begin{eqnarray} 
\chi_q (0,\mu)\!\!\!&=&\!\!\! \frac{\nu_q}{2\pi^2}\mu\sqrt{\mu^2-M_q^2}\
\theta(\mu-M_q)\ , \\
\chi_q (T, 0)\!\!\! &=&\!\!\!
\frac{\nu_q}{T}\!\int\!\!\!\frac{d^3k}{(2\pi)^3}\frac 1 {1+\cosh
  (E_q/T)}\ , \\ 
\label{eq:freesuscept}
\lim_{M_q\to 0}\chi_q (T, \mu)\!\!\! &=&\!\!\! \frac{\nu_q} 6 \left[ T^2 +
  \frac{3\mu^2}{\pi^2}\right]\equiv \chi_q^{\rm free}.
\end{eqnarray}

\begin{figure}[!htb]
  \centerline{\hbox{
      \includegraphics[width=0.6\columnwidth,angle=-90]{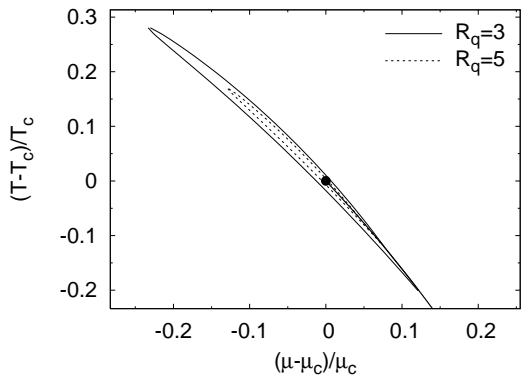}
    }}
\caption{\label{fig:q_sus_critreg_CEP} The contour regions for two
  different ratios of the quark number susceptibilities $R_q=\chi_q
  (T,\mu)/\chi_q^{\rm free}(T,\mu)$ ($R_q=3$ and $5$) in mean-field
  approximation around the CEP.}
\end{figure}

For an estimate of the critical region around the CEP we have
calculated the dimensionless ratio of the susceptibility and its
massless free quark gas limit (cf.~Eq.~(\ref{eq:freesuscept}))
\begin{equation}\label{eq:ratiochiq}
R_q (T,\mu)= \frac{\chi_q(T,\mu)}{\chi_q^{\rm free}(T,\mu)}\ .
\end{equation}

\noindent
Fig.~\ref{fig:q_sus_critreg_CEP} shows a contour plot for two fixed
ratios $R_q$ in the phase diagram near the CEP. A constituent quark
mass of roughly $300$ MeV has been chosen.  
The region of enhanced $\chi_q$ is elongated in the direction parallel 
to the first-order transition line.

In order to compare this behavior of the quark number susceptibility
around the CEP we repeat the previous analysis with the scalar
susceptibility. In general, static susceptibilities are obtained of
the dynamic response function $\chi_{ab}(\omega, \vec{q})$ in the
static and long wavelength limit
\begin{equation} 
  \chi_{ab} = - \frac{1}{V}\frac{\partial^2 \Omega }{\partial a
    \partial b} =\lim_{\vec q \to 0} \chi_{ab} (0, \vec{q})\ ,
\end{equation}
where $a,b$ denote external fields \cite{Fujii2004}. The scalar
susceptibility $\chi_\sigma$ corresponds to the zero-momentum
projection of the scalar propagator, which encodes all fluctuations of
the order parameter. The scalar susceptibility is related to the order
parameter by
\begin{equation}\label{eq:chisig}
  \chi_\sigma =\frac {\displaystyle \partial \vev{\bar qq} }
  {\displaystyle \partial m_q}=-\frac {\displaystyle \partial^2 \Omega }
  {\displaystyle \partial m_q^2}\ . 
\end{equation}
As a function of temperature or quark chemical potential the maximum
of $\chi_\sigma$ should coincide with the most rapid change in the
chiral order parameter. One easily verifies that the scalar
susceptibility is related to the sigma mass via $\chi_\sigma \sim
M_\sigma^{-2}$. 

Similar equations can also be formulated for the pseudoscalar
susceptibility $\chi_\pi$, which is related to the order parameter via
a Ward identity 
\begin{equation}\label{eq:chipi}
\chi_\pi = \vev{\bar qq}/m_q\ .
\end{equation}
In the chiral limit the divergence of the pseudoscalar (transverse)
susceptibility $\chi_\pi$ in the broken phase signals the appearance
of massless (Goldstone) modes.

\begin{figure}[!htb]
  \centerline{\hbox{
      \includegraphics[width=0.6\columnwidth,angle=-90]{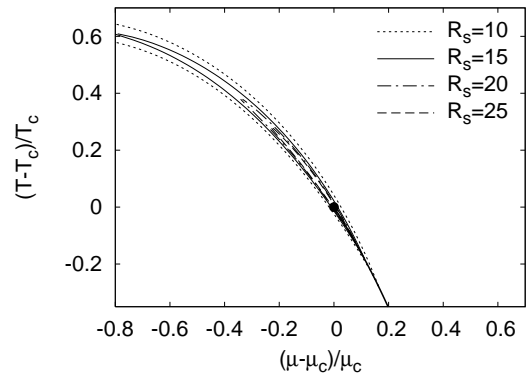}
    }}
\caption{\label{fig:crit_region_CEP} Same as
  Fig.~\ref{fig:q_sus_critreg_CEP} for four different ratios of the
  scalar susceptibilities $R_s=\chis (T,\mu)/\chis(0,0)$
  ($R_s =10,15,20,25$).}
\end{figure}

In Fig.~\ref{fig:crit_region_CEP} the contour region of the normalized 
scalar susceptibility 
\begin{equation}
R_s (T,\mu)= \frac{\chi_\sigma(T,\mu)}{\chi_\sigma(0,0)}
\end{equation}
is shown for four fixed ratios around the CEP. Again we observe an
elongated critical region in the phase diagram, where $\chi_\sigma$ is
enhanced in the direction parallel to the first-order transition line.
The deeper reason for this feature can be understood by a study of the
critical exponent of the susceptibility at the critical point which
will be done in the next Section.

\subsection{Critical exponents}

Since the quark-meson model in mean-field approximation does not
exhibit a tricritical point in the chiral limit we focus on the
critical behavior of the susceptibility in the vicinity of the CEP. At
that point the quark number susceptibility diverges with a certain
critical exponent. But a crucial observation is the following: In
general, the form of this divergence depends on the route by which one
approaches the critical point~\cite{Griffiths1970}.  For the path
asymptotically parallel to the first-order transition line the
divergence of the quark number susceptibility scales with an exponent
$\gamma_q$. In mean-field approximation one expects $\gamma_q=1$ for
this path. For any other path, not parallel to the first-order line,
the divergence scales with the exponent $\epsilon = 1-1/\delta$.
Thus, in mean-field approximation $\epsilon = 2/3$ because $\delta=3$
and therefore $\gamma_q > \epsilon$. This is the reason for the
elongated shape of the critical region in the phase diagram
(cf.~Figs.~\ref{fig:q_sus_critreg_CEP} and \ref{fig:crit_region_CEP})
and why $\chi_q$ is enhanced in the direction parallel to the
first-order transition line.

\begin{figure}[!htb]
  \centerline{\hbox{
      \includegraphics[width=0.6\columnwidth,angle=-90]{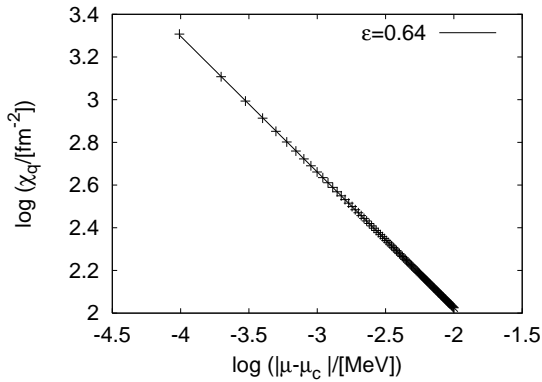}
    }}
\caption{\label{fig:scaling_chiq} The logarithm of the quark number
  susceptibility $\chi_q$ as a function of $\ln |\mu - \mu_c |$ at the
  CEP in mean-field approximation. The solid line is a linear fit. }
\end{figure}

In order to confirm this behavior we have calculated the critical
exponent of the quark number susceptibility $\chi_q$ numerically.  We
have used a path parallel to the $\mu$-axis in the ($T,\mu$)-plane
from lower $\mu$ towards the critical $\mu_c\sim 223$ MeV at fixed
temperature $T_c \sim 91.4$ MeV.  Fig.~\ref{fig:scaling_chiq} shows
the logarithm of $\chi_q$ as a function of $\mu$ close
to the CEP for a fixed constituent quark mass.  Using a linear
logarithmic fit we obtain
\begin{equation} 
\ln \chi_q = -\epsilon \ln |\mu -\mu_c | + r\ ,
\end{equation} 
where the term $r$ is independent of $\mu$. We observe scaling over
several orders of magnitude (only two orders are shown in the Figure)
and obtain $\epsilon = 0.64\pm 0.02$, which is in good agreement
with the mean-field prediction $\epsilon = 2/3$. The scaling close to
the critical point is also demonstrated in
Fig.~\ref{fig:scaling_chiq}. The onset of scaling is around $\ln |\mu
- \mu_c | < -1$ which is not shown in the Figure.

Furthermore, other critical exponents can also be calculated by using
susceptibilities.  Considering the ratio $R$ of the scalar and
pseudoscalar susceptibility at finite temperature it is possible to
calculate the critical exponent $\delta$~\cite{Kocic1993}. $R$ is a function 
of the reduced temperature $t$ and the bare quark mass $m_q$
\begin{equation} 
R(t,m_q )  = \chi_\pi^{-1} /  \chi_\sigma^{-1} = M_\pi^2 / M_\sigma^2\ .
\end{equation} 
At the critical point ($t=0$) the order parameter scales as $\vev
{\bar q q} \sim m_q^{1/\delta}$. Using
Eqs.~(\ref{eq:chisig}) and (\ref{eq:chipi}) one obtains 
\begin{equation} 
R(0,m_q) = 1/\delta\ ,
\end{equation}
which, at the critical point, is independent of $m_q$ and yields
$\delta=3$ in mean-field approximation. In order to obtain this value
it is necessary to determine the critical trajectory exactly at $t=0$
because there are two different limiting values for $R$ due to
spontaneous chiral symmetry breaking and its restoration.  In the
chiral limit the pions are Goldstone bosons in the broken phase
($t>0$) yielding $R(t>0,0) = 0$ and in the symmetric phase ($t<0$)
$M_\pi$ and $M_\sigma$ become degenerate at large temperatures giving
$R(t<0,0) = 1$.

\section{Proper-time RG approach} 
\label{Sec:PTRG}

In this section we study the phase diagram of the quark-meson model
within the PTRG approach in the chiral limit as well as for realistic
pion masses. In order to study the influence of fluctuations we will
repeat the calculations of all previous quantities  near the critical 
endpoint in the phase diagram.

As we have seen, the mean-field approximation fails to properly
describe the expected critical physics in the chiral limit (at least
for the parameter set chosen). We will show, that this is remedied in
the RG approach and a second-order transition in the chiral limit at
finite temperature and zero chemical potential is found. This
transition lies in the $O(4)$ universality class, as expected. For
nonzero chemical potential the second-order transition ends in a TCP
and becomes a smooth crossover for finite quark (pion) masses with a
CEP. Thus, by variation of the pion mass the relationship and the
correlations between the TCP and the CEP's can be studied. In
addition, the influence of fluctuations on the susceptibilities and
the critical region around the CEP can be assessed.

In order to include fluctuations we use the proper-time
renormalization group (PTRG) method~\cite{Liao1996, Schaefer1999,
  litim-2002-66, Litim:2002hj, litim-2002-65, Litim2006}. In the
vacuum, the PTRG flow equation for the scale-dependent effective
action $\Gamma_k [\Phi ]$ is governed by the selfconsistent equation
\begin{equation*} 
\partial_t \Gamma_k [\Phi] = -\frac 1 2 \int\limits_0^\infty \frac
{d\tau}\tau \left[ \partial_t f_k (\tau k^2)\right] tr
\exp\left(-\tau \Gamma^{(2)}_k [\Phi ]\right)\ , 
\end{equation*}
where $\Gamma^{(2)}_k$ represents the full inverse propagator and is
given by a second functional derivative w.r.t. the field components
$\Phi$. The smearing function is labeled by $f_k (\tau k^2)$ and $tr$
denotes a four-dimensional momentum integration and a trace over all
given inner spaces (e.g. Dirac, color and/or flavor-space).  Details
and the generalization to finite temperature and
chemical potential can be found e.g.~in
Refs.~\cite{Schaefer2005,Schaefer1999}.

For the quark-meson model (\ref{eq:qmmodel}) the resulting flow
equation for the scale-dependent grand canonical potential $\Omega_k
(T,\mu)$ is given by
\begin{eqnarray} 
\label{eq:fullRG}
\partial_t \Omega_k(T,\mu) = \frac {k^5} {12\pi^2} 
\left[  \frac 3 {E_\pi} \coth \!\!\left(\frac {E_\pi}{2T} \right)\!+\!
    \frac 1  {E_\sigma} \coth \!\!\left(\frac
      {E_\sigma}{2T} \right) \right.&&\nonumber \\
\left.- \frac {2 N_c N_f}{E_q}\left\{ \tanh\!\!  \left(\frac 
     {E_q -\mu} {2T}\right) +\tanh\!\! \left(\frac
     {E_q +\mu} {2T}\right)\right\}\right] ,\hspace*{\fill}\  &&
\end{eqnarray} 
with the pion- $E_\pi = \sqrt{k^2 + 2 \Omega'_k}$, the $\sigma$-meson
$E_\sigma = \sqrt{k^2 + 2 \Omega'_k + 4\phi^2 \Omega''_k}$ and quark
energies $E_q = \sqrt{k^2 + g^2 \phi^2}$. The primed potential denotes
the $\phi^2$-derivative of the potential, i.e., $\Omega'_k := \partial
\Omega_k / \partial \phi^2$ and correspondingly the higher
derivatives.

We solve the flow Eq.~(\ref{eq:fullRG}) by discretizing the generally
unknown potential $\Omega_k$ on a $\phi^2$-grid. For details
concerning the numerical implementations we refer the interested
reader to Refs.~\cite{Bohr2001,Schaefer2005} and references therein.

For the results presented below we have chosen an UV cutoff $\Lambda =
500$ MeV and have fixed the quartic coupling to $\lambda =11.5$ in
order to reproduce a pion decay constant $f_\pi \sim 87$ MeV in the
chiral limit.  The running of the Yukawa coupling $g$ is neglected
here and fixed to $g=4.2$ in order to reproduce a constituent quark
mass of the order of $370$ MeV. Note, that this value of the quark
mass is roughly 100 MeV larger than the one used in the previous
work~\cite{Schaefer2005}.  As a consequence, the tricritical point in
the phase diagram moves to larger temperatures and smaller chemical
potentials.

\subsection{The phase diagram}

\begin{figure}[!htb]
  \centerline{\hbox{
      \includegraphics[width=0.6\columnwidth,angle=-90]{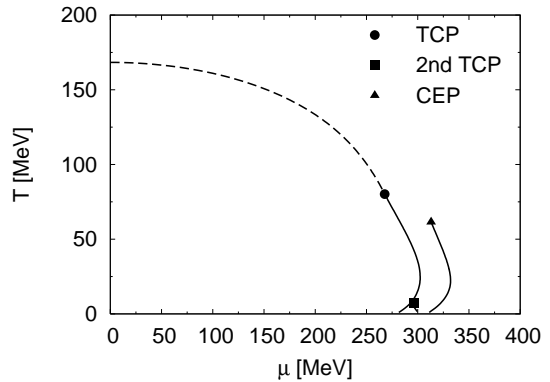}
    }}
  \caption{\label{fig:rgphasediagram} Two phase diagrams of the linear
    $N_f=2$ quark-meson model: One for physical pion masses (right
    solid line which ends in the CEP) and another one for the chiral
    limit. Solid lines denote first-order and dashed lines 
    second-order transition lines (see text for details).}
\end{figure}

In the present work we have also generalized previous results
in the chiral limit~\cite{Schaefer2005} to finite
pion masses. The resulting phase diagrams for the chiral limit and for
physical pion masses $M_\pi \sim 130$ MeV are both shown in
Fig.~\ref{fig:rgphasediagram}. In the chiral limit a second-order
phase transition (dashed line) belonging to the $O(4)$-universality
class is found. For zero chemical potential we find a critical
temperature $T_c \sim 170$ MeV for two massless quark flavors, in good
agreement with lattice simulations~\cite{Karsch2001}. For increasing
chemical potential the second-order transition line ends in a TCP
(bullet).

For our choice of parameters the location of the TCP is at $T^t_c \sim
80$ MeV and $\mu^t_c = 268$ MeV. For temperatures below the TCP the
phase transition changes initially to a first-order transition (solid
line). For temperatures below $T_s \sim 10$ MeV two phase transitions
with a further tricritical point (labeled as '2nd TCP' in the Figure)
emerge as already described in Ref.~\cite{Schaefer2005}. A larger
constituent quark mass pushes the location of the TCP towards the
temperature axis and the location of the splitting point down towards
the chemical potential axis. All qualitative properties of the two
phase transitions below the splitting point survive. Only the area
bounded by the transition lines is reduced for increasing quark masses
(cf.~phase diagram in Ref.~\cite{Schaefer2005}).

As the pions become massive the TCP turns into a CEP. For each value
of the pion mass there is a corresponding CEP. In an extended,
three-dimensional ($T$, $\mu$, $m_\pi$) phase diagram these points
arrange into a critical line (the 'wing critical line'). It is
expected that the static critical behavior of this line falls into the
universality class of the Ising model in three dimensions
corresponding to the one-component $\phi^4$-theory in 3D. For the
coupling constants listed above $M_\pi \sim 123$ MeV and $M_q \sim
390$ MeV in the vacuum and the location of the CEP is at $T_c^c \sim
61.5$ MeV and $\mu_c^c \sim 313.1$ MeV. A better fit to the physical
pion mass can be accomplished by a fine tuning of the initial coupling
constants. Compared to recent lattice and other model studies
(cf.~Fig.~6 in Ref.~\cite{Stephanov2005}) the location of the TCP and
consequently of the CEP is at lower temperatures due to omission of
other degrees of freedom in the quark-meson model.

For any finite value of the symmetry breaking parameter $c$ in the
potential (\ref{eq:pot}) the critical $O(4)$-line is immediately
washed out and turns into a smooth crossover line which is not visible
in the phase diagram. For temperatures below the CEP a first-order
curved transition line is found which persists down to the $\mu$-axis.
In the chiral limit below the splitting point, the right second-order
transition turns into a crossover for finite quark masses and is again
not visible in the phase diagram anymore. Analogously, the second
tricritical point, if it exists, should turn into a critical point.
Some remnants of this critical point can indeed be seen in the vacuum
expectation value and meson masses. But a detailed analysis of this
point is postponed to a future work. In the following we focus of the
region around the TCP and CEP.

\subsection{In-medium meson masses}

The scale-dependent in-medium sigma and pion masses are determined by
the curvature of the potential $\Omega$ at the global minimum $\phi_0$
via the relations
\begin{equation} 
M_{\pi,k}^2 = \left.2\Omega'_k\right|_{\phi=\phi_0}\ ,\quad
M_{\sigma,k}^2 = \left.(2\Omega'_k+4\phi^2\Omega''_k
  )\right|_{\phi=\phi_0}\ .
\end{equation}

\begin{figure}[!htb]
  \centerline{\hbox{
      \includegraphics[width=0.6\columnwidth,angle=-90]{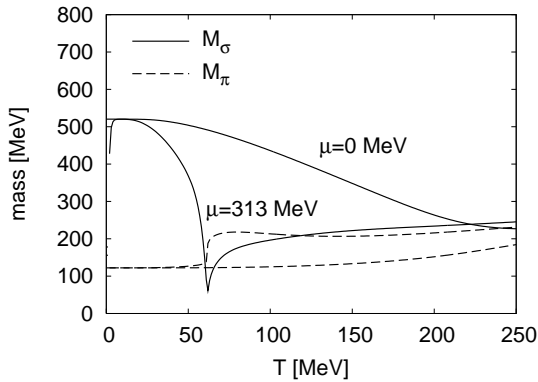}
    }}
  \caption{\label{fig:RG_masses_T} The pion and sigma masses as a
    function of temperature for two different quark chemical
    potentials ($\mu = 0$ and $\mu = \mu^c_c \sim 313$ MeV). Near the
    CEP (located at $\mu^c_c$ and $T^c_c \sim 61.5$ MeV) $M_\sigma$
    vanishes.}
\end{figure}

In Fig.~\ref{fig:RG_masses_T} the meson masses are shown as a function
of temperature for two different chemical potentials similar to
Fig.~\ref{fig:masses_T}. For $\mu =0$ the sigma mass drops relatively
quickly with temperature while the pion mass are almost constant in
the chirally broken phase. This behavior is very similar to a PTRG
analysis with another smearing function and a truncation of the meson
potential (cf.~Fig.~14 in \cite{Schaefer1999}). For $\mu=\mu^c_c\sim
313$ MeV and at $T=0$ the sigma mass is almost degenerate with the
pion mass $M_\pi \sim 150$ MeV (see Fig.~\ref{fig:RG_masses_T}). For
increasing temperatures, the sigma mass jumps to almost its vacuum
value as it crosses the first-order phase transition. In the vicinity
of the CEP around $T\sim 60$ MeV the sigma mass drops below the pion
mass and vanishes at the critical point. At this point the potential
in radial direction around the global minimum becomes flat. As in the
mean-field approximation the slope of the sigma mass as function of
temperature (or of the chemical potential) at the CEP tends to
infinity.

\begin{figure}[!htb]
  \centerline{\hbox{
      \includegraphics[width=0.6\columnwidth,angle=-90]{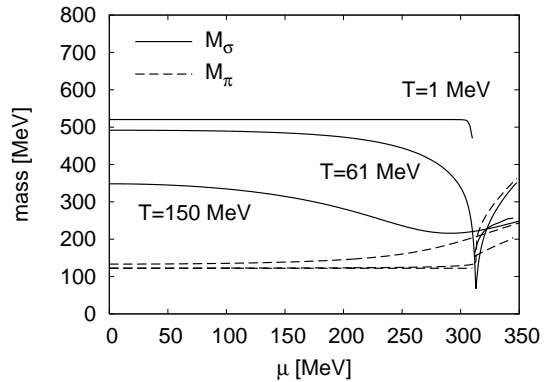}
    }}
  \caption{\label{fig:RG_masses_mu} Similar to
    Fig.~\ref{fig:RG_masses_T}: The pion and sigma masses as a
    function of chemical potential for three different temperatures
    ($T=1$, $T=T^c_c \sim 61$ and $T=150$ MeV).}
\end{figure}

In Fig.~\ref{fig:RG_masses_mu} the meson masses parallel to the
$\mu$-axis for three different temperature slices ($T=1$, $T=T^c_c$
and $T=150$ MeV) in the phase diagram are shown. For $T=1$ MeV the
sigma mass is constant and slightly before the first-order transition
a melting of the mass takes place.

Due to the curved first-order transition line the critical chemical
potential at the CEP around $T\sim 61$ MeV coincides with the critical
chemical potential at $T=1$ MeV.

One important observation is that the sigma mass drops to zero around
the CEP much faster if fluctuations are included.

\subsection{Susceptibilities}

In the vicinity of critical phenomena, fluctuations which are
neglected in a mean-field approximation become more and more
important. In order to study the modifications, induced by the
fluctuations, we have recalculated the quark number density and
susceptibilities within the RG method. It should be recapitulated
that, in contrast to the mean-field approximation, this method yields
a TCP in the phase diagram for the chiral limit.

\begin{figure}[!htb]
  \centerline{\hbox{
      \includegraphics[width=0.6\columnwidth,angle=-90]{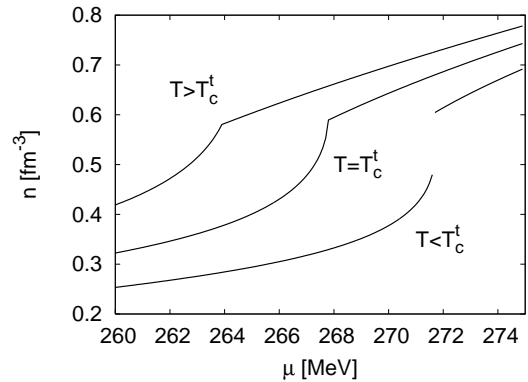}
    }}
  \caption{\label{fig:quarknumber_TCP} The quark number density around
    the TCP. The three temperatures are $T^t_c = 80.2$ MeV and $T =
    T^t_c \pm 5$ MeV. }
\end{figure}

In the following we begin our investigations with the chiral limit. In
Fig.~\ref{fig:quarknumber_TCP} the density $n$ is shown as function of
$\mu$ for three different temperatures around the TCP.

\begin{figure}[!htb]
  \centerline{\hbox{
      \includegraphics[width=0.6\columnwidth,angle=-90]{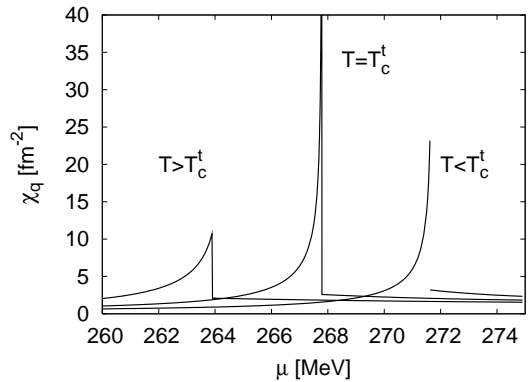}
    }}
\caption{\label{fig:q_suscept_TCP} The quark number susceptibility around
  the TCP. The three temperatures are the same as in
  Fig.~\ref{fig:quarknumber_TCP}. }
\end{figure}

For temperatures below the tricritical $T^t_c$ $n$ jumps due to the
first-order transition. Above the tricritical temperature there is a
kink in the density due to the second-order nature of the transition.
In the symmetric phase $n$ grows almost linearly. At the TCP the slope
of the density when approached from the broken phase, diverges thus
yielding a divergent susceptibility. This is demonstrated in
Fig.~\ref{fig:q_suscept_TCP} where the chemical potential dependence
of the quark number susceptibility is shown for three different fixed
temperatures around the TCP.

In the chiral limit, the susceptibility always jumps across the first-
or the second-order transition. In Fig.~\ref{fig:q_suscept_TCP} the
second-order transition jump of $\chi_q$ is drawn with solid lines.
Far below the chiral phase transition $\chi_q$ is suppressed. In the
restored phase $\chi_q$ tends towards the value of the massless free
quark gas, $\chi_q^{\rm free}$, Eq.~(\ref{eq:freesuscept}).

\begin{figure}[!htb]
  \centerline{\hbox{
      \includegraphics[width=0.6\columnwidth,angle=-90]{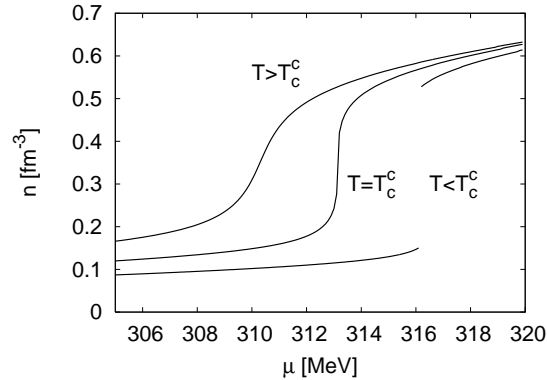}
    }}
  \caption{\label{fig:quarknumber_CEP} The quark number density around
    the CEP. The three temperatures are $T^c_c = 61.5$ MeV and $T =
    T^c_c \pm 5$ MeV. }
\end{figure}

When we now leave the chiral limit and investigate the phase diagram
around the CEP, the critical behavior of the transition changes. In
contrast to the behavior of the quark number density around the TCP it
does not have any kink at finite quark masses
(cf.~Fig.~\ref{fig:quarknumber_CEP}).

\begin{figure}[!htb]
  \centerline{\hbox{
      \includegraphics[width=0.6\columnwidth,angle=-90]{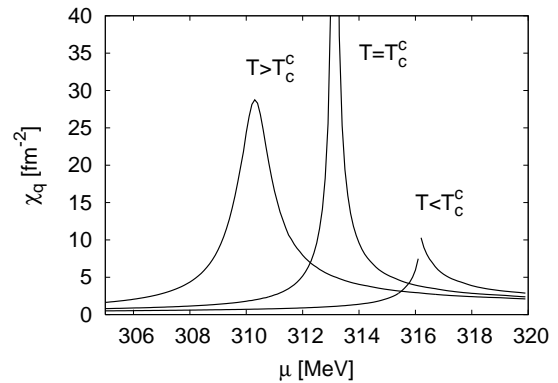}
    }}
\caption{\label{fig:suscept_CEP} The quark number susceptibility around
  the CEP as a function of chemical potential for three fixed
  temperatures. The three temperatures are $T_c^c = 61.5$ MeV and $T =
  T_c^c \pm 5$ MeV.}
\end{figure}

For temperatures below the critical $T^c_c$, the density jumps again
at the phase transition due to its first-order character. But above
the CEP the discontinuity vanishes at the transition and the system
runs through a smooth crossover. For temperatures above $T^c_c$, the
slope of the density is always finite, except exactly at $T^c_c$ where
it diverges yielding a divergent quark number susceptibility at this
point. Thus, the divergence of $\chi_q$ survives even at finite quark
masses. The finite slope results in a finite peak of $\chi_q$ which is
shown in Fig.~\ref{fig:suscept_CEP} where the quark number
susceptibility for three different temperatures around the CEP,
similar to Fig.~\ref{fig:q_suscept_TCP}, is displayed.

Below the chiral transition, the susceptibility is again suppressed
and in the restored phase it tends towards $\chi_q^{\rm free}$ of a
massless free quark gas. Only at the first-order transition there is a
discontinuity. In contrast to the chiral limit, the quark
susceptibility is a smooth continuous function for temperatures above
the CEP.

On the lattice a similar behavior of the quark susceptibility is
seen~\cite{Allton2005,Gottlieb1987}. At $\mu=0$, $\chi_q$ increases
smoothly near the critical temperature.  At finite chemical potential
it develops a pronounced dip but with increased error bars around the
transition temperature.  On the other side, a significant peak in the
isovector channel is not seen on the lattice. Thus, this indicates
that only one scalar degree of freedom becomes massless at this point.
Also universality arguments predict that the susceptibility diverges
at both the TCP and CEP with certain but different critical exponents.
At the TCP $\chi_q$ should show a power-law behavior with a critical
exponent $\gamma_q$. Since the TCP belongs to a Gaussian fixed point
mean-field exponents are expected.

In the chiral limit the susceptibility $\chi_q$ always has a
discontinuous jump across the critical $O(4)$-transition line.
$\chi_q$ is larger below the $O(4)$-line (in the chirally broken
phase) than in the restored phase. At any finite $\mu$ the
susceptibility has a cusp at the critical $O(4)$-transition line. The
peak of this cusp on the critical line as approached from the broken
phase becomes higher and higher as we increase $\mu$. Finally, it
diverges exactly at the TCP, while the potential and thus the pressure
will stay finite.  Only at $\mu=0$, the quark number susceptibility is
a continuous function across the phase boundary.

On the other side, as already mentioned, in the chirally symmetric
phase, $\chi_q$ tends towards the value of the massless free quark-gas
susceptibility $\chi_q^{\rm free}$. The deviations are caused by the
residual meson interaction in this phase.

\subsection{Critical exponents}

For finite quark masses it is expected that the universality class of
the critical points changes. In order to determine the universality
class of the TCP and CEP we have also calculated the corresponding
critical exponents of the quark number susceptibility. In determing
the critical exponent we proceed in exactly the same way as described
in the mean-field section. The calculation of the first and second
derivative of the potential for the quark number density and
susceptibility is again determined numerically with an adaptive step
size algorithm in order to minimize rounding and truncation errors.

We again begin the analysis with the chiral limit. We have used a path
parallel to the $\mu$-axis in the $(T,\mu)$-plane from lower $\mu$
towards the TCP. Note, that in the $(T,\mu)$-plane the first-order
transition line and the critical $O(4)$-line at the TCP are
asymptotically parallel and are not parallel to either the $T$- or
$\mu$-axis.

\begin{figure}[!htb]
  \centerline{\hbox{
      \includegraphics[width=0.6\columnwidth,angle=-90]{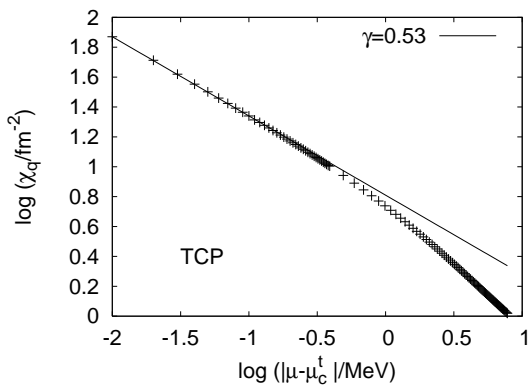}
    }}
\caption{\label{fig:tcp_suscept_exponent} The logarithm of the
  quark number susceptibility around the TCP as a function of the
  logarithm of the chemical potential in the vicinity of the TCP.
  The solid line is a linear fit.}
\end{figure}

Using a linear logarithmic fit of $\chi_q$ in the vicinity of the TCP
we obtain the critical exponent $\gamma_q = 0.53$ which is shown in
Fig.~\ref{fig:tcp_suscept_exponent}. The region where the scaling
starts is rather small $|\mu~-~\mu^t_c| < 10^{-0.5} \sim 0.3$ MeV.  This
exponent is consistent with mean-field theory which predicts a
critical exponent $\gamma_q = 1/2$. This is also expected due to the
Gaussian fixed point structure of the TCP.

\begin{figure}[!htb]
  \centerline{\hbox{
      \includegraphics[width=0.6\columnwidth,angle=-90]{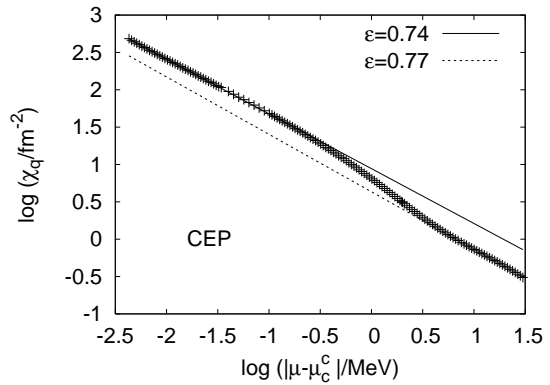}
    }}
\caption{\label{fig:q_suscept_CEP_scaling} Same as
  Fig.~\ref{fig:tcp_suscept_exponent} for the CEP.}
\end{figure}

Leaving the chiral limit, the universality class of the CEP is that of
the three-dimensional Ising model. Using again a path parallel to the
$\mu$-axis from lower $\mu$ towards the CEP we repeat the calculation
of the critical exponent. In Fig.~\ref{fig:q_suscept_CEP_scaling}
$\chi_q$ is shown versus the distance from the CEP. It is interesting
to see, that there seems to be two different scaling regimes, which we
indicated with two linear fit functions in the figure.

The slope for the data points changes between $10^{-0.5}$ MeV $ < |\mu
- \mu^c_c| < 10^{0.5}$ MeV. We have fitted the data for $ |\mu -
\mu^c_c| < 10^{-0.5}$ MeV and $> 10^{0.5}$ MeV separately and obtained
the critical exponent $\epsilon \sim 0.74$ (solid line) for $ |\mu -
\mu^c_c| < 10^{-0.5} \sim 0.3$ MeV. For $|\mu - \mu^c_c| > 10^{0.5}$
MeV we also see a linear behavior for several orders of magnitude. If
there is a proper scaling behavior of the susceptibility in this
region, the slope would be consistent with an exponent of $\epsilon
\sim 0.77$ (dashed line). This change of the exponents could be
interpreted as a crossover of different universality
classes~\cite{Gebhardt1980}. One possibility for this crossover
phenomenon is that the CEP is affected by the TCP. In
Ref.~\cite{Hatta2003} a similar crossover phenomenon between different
universality classes in the framework of the CJT potential for QCD in
improved-ladder approximation is seen: As they approach the CEP for
realistic quark masses ($m_q \sim 5$ MeV), the critical exponent
change gradually from those of the tricritical point to those of the
3D Ising model via those of the CEP in the mean-field approximation.
In mean-field theory one expects $\epsilon \equiv \gamma/\beta \delta
= 2/3$. They find a crossover from the nontrivial critical exponent
$\epsilon = 0.57\pm0.01$ to the mean-field exponent $\epsilon =
0.68\pm0.02$. Contrary to this work, we find a crossover from
$\epsilon = 0.77$ to $\epsilon = 0.74$. Thus, at the CEP the
susceptibility diverges with the critical exponent $\epsilon \sim
0.74$. This exponent is consistent with the one of the 3D Ising
universality class $\epsilon = 0.78$ and is definitely different from
the mean-field value $\epsilon = 2/3$. Note also, that the mean-field
exponent of a bicritical point (CEP) are in general different from
those of a tricritical point. In order to complete the analysis here
we summarize the critical exponents of different approaches in
Tab.~\ref{tab:critexp}.

\begin{table}[h!]
  \begin{tabular}{r||@{\hspace{1mm}}c@{\hspace{1mm}}|@{\hspace{1mm}}
      c@{\hspace{1mm}}|@{\hspace{1mm}}c@{\hspace{1mm}}|@{\hspace{1mm}}
      c@{\hspace{1mm}}|@{\hspace{1mm}}c@{\hspace{1mm}}|@{\hspace{1mm}}
      c@{\hspace{1mm}}||l}
    & $\alpha$ & $\beta$ & $\gamma$ & $\delta$ & $\nu$ & $\eta$ & Ref.\\
    \hline
    \hline
    mean-field  & 0 & 1/2 & 1 & 3 & 1/2 & 0 & \cite{Gebhardt1980} \\
    \hline
    $\epsilon$-exp. (5 loops) & 0.11 & 0.327 & 1.24 & 4.80 & 0.631 &
    0.04 & \cite{Zinn-Justin1999} \\
    \hline
    num. sim. & 0.12 & 0.31 & 1.25 & 5.20 & 0.64 & 0.06 & 
    \cite{Guida1997} \\
    \hline
  \end{tabular}
  \caption{\label{tab:critexp} Critical exponents for the 3D Ising model.}
\end{table}

It is also interesting that the value $|\mu - \mu^c_c| \sim 0.5$ MeV
where the slope of the data points changes to a mean-field scaling in
Ref.~\cite{Hatta2003} is comparable to our value.

\subsection{The critical region}

\begin{figure}[!htb]
  \centerline{\hbox{
      \includegraphics[width=0.6\columnwidth,angle=-90]{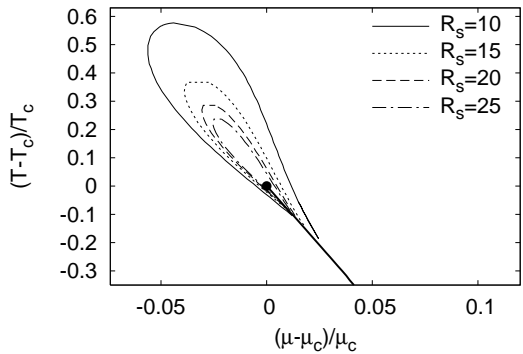}
    }}
\caption{\label{fig:cr_CEP} The contour regions for four different
  ratios of the scalar susceptibilities $\chis (T,\mu)/\chis(0,0)$
  ($R_s = 10,15,20,25$) around the CEP in the phase diagram.}
\end{figure}

We are now able to compare the impact of the quantum and thermal
fluctuations on the shape of the region around the critical points in
the phase diagram. We repeat the calculation of the contour plot of
Figs.~\ref{fig:q_sus_critreg_CEP} and \ref{fig:crit_region_CEP} in the
framework of the RG approach. The result for the scalar susceptibility
around the CEP for the same four ratios $R_s$ as in the mean-field
approximation is shown in Fig.~\ref{fig:cr_CEP}.

As in the mean-field case the region is elongated in the direction of
the first-order transition line, but it is now much more compressed.
For example, choosing a ratio of $R_s = 15$, the corresponding
susceptibility covers an interval from $-0.04$ to $0.02$ in the
reduced chemical potential direction and an interval from $-0.1$ to
$0.4$ in the temperature direction. In the mean-field case the same
ratio covers an interval from $-0.8$ to $0.1$ in the $\mu$ direction
and from $-0.15$ to $0.6$ in the $T$ direction. While the interval in
the temperature direction is comparable in both cases, the effect in
the chemical potential direction is dramatic. In the RG calculation
the interval is shrunken by almost one order of magnitude, despite the
fact that the corresponding critical exponents are quite similar!

\begin{figure}[!htb]
  \centerline{\hbox{
      \includegraphics[width=0.6\columnwidth,angle=-90]{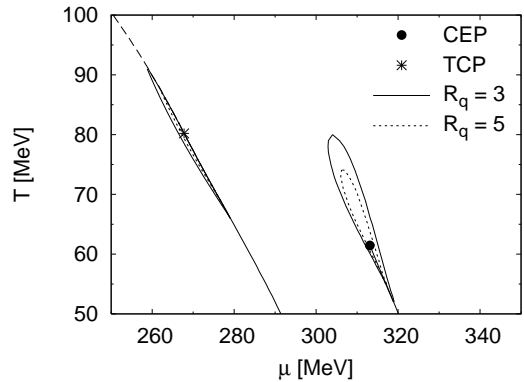}
    }}
\caption{\label{fig:crit_reg_CEP} The contour regions for two different ratios
  of the quark number susceptibilities $R_q = \chi_q
  (T,\mu)/\chi_q^{\rm free}(T,\mu)$ around the CEP and TCP in the
  phase diagram.}
\end{figure}

An similar result is obtained for the critical region of the
quark number fluctuations. In order to compare the critical region
around the CEP with the one around the TCP we show the quark number
susceptibility in Fig.~\ref{fig:crit_reg_CEP} in a larger sector of
the phase diagram containing both critical points.

The CEP is far away from the TCP at larger values of the chemical
potential and smaller temperature as expected. Due to the sharp
transition lines in the chiral limit the critical region around the
TCP is chop off in the chirally symmetric phase. Whether there is a
robust effect of the TCP on the CEP as stated in Ref.~\cite{Hatta2003}
is not seen in this work. But this important issue including the quark
mass dependence of the susceptibilities is postponed to a further
analysis.

\section{Summary and Conclusion}
\label{Sec:sum}

Using a Wilsonian renormalization group approach, we have analyzed the
phase diagram of hadronic matter in the two-flavor quark-meson model.
This model captures essential features of QCD, such as the spontaneous
breaking of chiral symmetry in the vacuum and can therefore yield
valuable insight into the critical behavior, associated with chiral
symmetry. Of special importance is the emergence of a CEP which is
intensely discussed at present, in connection with fluctuation signals
in heavy-ion collisions. Here the size of the critical region around
the CEP is of special importance. Most studies of this issue,
available in the literature, have been performed in the mean-field
approximation, which neglects thermal and quantum fluctuations. These
can be assessed, however, in the RG approach which is able to
correctly predict critical exponents in the vicinity of critical
points of the phase diagram.

The main results of this work can be summarized as follows: 

\begin{enumerate}
\item From universality arguments it is expected that the quark-meson
  model (and most likely QCD) has a TCP in the chiral limit. For the
  parameter set chosen in this paper, a mean-field calculation is not
  able to find such a point, while the RG predicts its existence. The
  expected critical behavior is also reproduced. Due to the Gaussian
  fixed-point structure at the TCP mean-field exponents are expected
  what we could verify.

\item When effects of finite current quark masses (finite pion masses)
  are included, a CEP emerges in both the mean-field and the RG
  calculation. By analyzing the scalar- and quark number
  susceptibilities we find that the RG calculation yields non-trivial
  critical exponents, consistent with the expected 3D Ising
  universality class. Close to the CEP our exponent is consistent with
  the Ising class but we also see a novel crossover phenomenon.
 
\item As a consequence of fluctuations, the size of the critical
  region around the CEP is substantially reduced, as compared to
  mean-field results. This is particularly true in the
  $\mu$-direction, where a shrinkage by almost one order of magnitude
  is observed. This may have consequences for the experimental
  localization of the CEP in the phase diagram since it further
  complicates its detection through event-by-event fluctuations.

\end{enumerate}

The success of the RG approach in predicting the expected critical
behavior of the thermodynamics in the quark-meson model encourages us
to pursue the issue of the existence of a CEP in the phase diagram of
strongly interacting matter and its location in the ($T,\mu$)-plane
further. It is known from lattice studies that the strange quark plays
a major role. We therefore intend to extend the analysis to three
quark flavors~\cite{Schaefer2006}. One of the drawbacks of the
quark-meson model is the lack of explicit gluonic degrees of freedom,
which are known to play a big role in the thermodynamics of QCD and
are associated with confinement aspects of the theory. One possibility
to incorporate such effects is the coupling of the quark-meson model
to Polyakov-loop model~\cite{Dumitru2002,Dumitru2004}, as put forward
in~\cite{Ratti:2005jh}. Work in this direction is in
progress~\cite{Pawlowski2006}.

\section*{Acknowledgment}

One of the authors (BJS) thanks for illuminating discussions with
M.~Buballa, B.~Drossel and E.~Shuryak. This work is supported in part
by the Helmholtz association (Virtual Theory Institute VH-VI-041) and
by the BMBF grant 06DA116.


\end{document}